\documentclass[final, 5p, times, twocolumn]{elsarticle}
\journal{Physics Letters B}

\usepackage{color}

\newcommand{\apj}{Astrophys~J.}
\newcommand{\apjl}{Astrophys~J.~Lett.}
\newcommand{\apjs}{Astrophys~J.~Sup.}
\newcommand{\prc}{Phys.~Rev.~C}
\newcommand{\physrep}{Phys.~Rep.}
\newcommand{\aap}{Phys.~Rep.}
\newcommand{\araa}{ARAA}
\newcommand{\mnras}{MNRAS}

\newcommand{\FF}{{{\it FF}}}

\begin{document}

\begin{frontmatter}

\title{Impact of the first-forbidden $\beta$~decay on the production of $A \sim 195$ {\it r}-process peak}

\author{Nobuya~Nishimura$^a$, Zsolt~Podoly\'ak$^b$, Dong-Liang~Fang$^c$, Toshio~Suzuki$^{d, e}$}
\address{$^a$ Astrophysics Group, Keele University, Keele ST5 5BG, UK; n.nishimura@keele.ac.uk}
\address{$^b$ Department of Physics, University of Surrey, Guildford GU2 7XH, UK}
\address{$^c$ College of Physics, Jilin University, Changchun, Jilin 130012, China}
\address{$^d$ Department of Physics, Nihon University, Sakurajosui, Tokyo 156-8550, Japan}
\address{$^{e}$ Division of Theoretical Astronomy, National Astronomical Observatory of Japan, Mitaka, Tokyo 181-8588, Japan}

\begin{abstract}
We investigated the effects of first-forbidden transitions in $\beta$ decays
on the production of the {\it r}-process $A \sim 195$ peak.
The theoretical calculated $\beta$-decay rates with $\beta$-delayed neutron emission
were examined using several astrophysical conditions.
As the {\FF} decay is dominant in $N \sim 126$ neutron-rich nuclei,
their inclusion shortens $\beta$-decay lifetimes and shifts the abundance peak towards higher masses.
Additionally, the inclusion of the $\beta$-delayed neutron emission results in a wider abundance peak,
and smoothens the mass distribution by removing the odd-even mass staggering.
The effects are commonly seen in the results of all adopted astrophysical models.
Nevertheless there are quantitative differences,
indicating that remaining uncertainty in the determination of half-lives for $N=126$ nuclei
is still significant in order to determine the production of the {\it r}-process peak.
\end{abstract}


\end{frontmatter}

\section{Introduction}
\label{intro}

The rapid neutron-capture process ({\it r}~process) is one of the major
nucleosynthesis processes \cite{1957RvMP...29..547B, 1997RvMP...69..995W},
producing nuclei heavier than iron, which include rare-earth elements and actinides.
The {\it r}~process is considered to take place
in explosive neutron-rich environments in the universe
\citep[][]{2007PhR...450...97A, 2011PrPNP..66..346T},
e.g., core-collapse supernovae (CC-SNe), which are the formation process of a neutron star (NS),
and/or NS--NS mergers, coalescences of binary NSs.
The realistic astrophysical scenario, however, is still unsettled
even based on recent sophisticated multi-dimensional hydrodynamical simulations.
On the other hand, the basic nucleosynthesis mechanism and the required physical conditions
are relatively well understood.
The {\it r}~process consists of multiple neutron captures (n,$\gamma$)
competing with the photo-disintegration ($\gamma$,n) and $\beta$ decay,
so that nucleosynthetic path is laid down on very neutron-rich region far from the stability.
The {\it r}~process has been extensively investigated based on nuclear reaction network calculations
using simplified astrophysical models
\cite{1992ApJ...395..202W, 1992ApJ...399..656M}.
For a given astrophysical (hydrodynamical) condition,
the final abundances are uniquely determined
by the nuclear physics properties,
i.e., masses, decay half-lives, $Q$-values and (n,$\gamma$) and ($\gamma$,n) cross sections
(see \citep{2016PrPNP..86...86M} and references therein).

The observed {\it r}-process abundance pattern\footnote{ 
Several {\it r}-process-enriched metal-poor stars show solar-like {\it r}-process abundance distribution
in the $Z > 60$ ($A>120$) region \cite{2008ARA&A..46..241S}.
In this study, we can safely assume the solar abundances as a typical (``universal'') {\it r}-process pattern,
because we mostly focus on heavy {\it r}-process nuclei ($A>150$).}
has peaks around $A \sim 130$ and $195$,
which originate from waiting point nuclei around $N=82$ and $126$ neutron magic numbers
along the {\it r}-process nucleosynthesis path.
For nuclei with magic neutron numbers, the neutron capture becomes inhibited
and the nucleosynthesis flow is influenced by the $\beta$~decay.
In addition, after terminating the neutron capture,
the final abundances are determined
by the $\beta$~decay with emission of neutrons.
Thus the $\beta$-decay properties of $N=82$ and $126$ nuclei,
which are half-lives and corrections from $\beta$-delayed neutron emission,
are important for discussing final abundances \cite{2003RvMP...75..819L, 2012EPJA...48..184B}.
Despite the importance, half-lives were measured for only a portion of neutron-rich nuclei
and most of the waiting point nuclei cannot be studied experimentally at the moment.
In the recent years, a lot of progress was obtained for nuclei $N\sim82$
at radioactive beam facilities such as RIBF at RIKEN
\cite{2011PhRvL.106e2502N, 2012PhRvC..85d8801N, 2015PhRvL.114s2501L}.
However, experimental information is not available in $N \sim 126$ region
(measurements are performed only near the line of stability \cite{2014PhRvL.113b2702M}),
so that nucleosynthesis calculations have to rely on theoretical predictions.

Theoretical prediction of $\beta$-decay half-lives has been traditionally performed
by considering only allowed transitions.
The allowed $\beta$-decay transitions, which conserve parity
and the change of angular momentum is limited to $0$ or $1$,
are dominant for the majority of nuclei.
This is valid for the $N=50$ and $82$ neutron closed shell nuclei,
i.e., the previous calculations are reliable.
However, in cases when a large contributions from first-forbidden (FF) $\beta$~decay are expected,
these calculations severely overestimate the lifetimes.
This is significant for nuclei around the $N=126$ waiting point, where the {\FF} decay can be dominant.
In this mass region, the {\FF} decay, based on the $\nu i_{13/2}$ $\rightarrow$ $\pi h_{11/2}$ transition,
competes with the allowed decays due to 
the $\nu h_{9/2}$ $\rightarrow$ $\pi h_{11/2}$ transition.
This is because the {\FF} decay is energetically favoured
(the $h_{9/2}$ neutron is deep in the shell, while the $i_{13/2}$ is much closer to the Fermi level).

Currently the most widely adopted theoretical $\beta$-decay half-lives
are based on the finite-range droplet model (FRDM)
\cite{1995ADNDT..59..185M}.
They originally included only allowed transitions \cite{1997ADNDT..66..131M},
with {\it FF} transitions incorporated later using the gross theory \cite{2003PhRvC..67e5802M}.
More recently, several calculations take into account first-forbidden $\beta$ decays
by the extended quasiparticle random phase approximation (QRPA)
\cite{2006NuPhA.777..645B, 2011PAN....74.1435B, 2013PhRvC..88c4304F, 2014arXiv1405.0210C, 2015ApJ...808...30E}
and shell model approaches \cite{2012PhRvC..85a5802S, 2013PhRvC..87b5803Z}.
These studies found a significant contribution, in the range of $20$ -- $80$\%,
of the {\FF} transition to the $\beta$ decay for $N = 126$ waiting point nuclei.
These calculations also predict a large $\beta$-delayed neutron emission probability
for the $N \sim 126$ {\it r}-process path nuclei,
which have a high $Q_\beta$ value and low neutron separation energy.
Thus, the $\beta$~decay may populate states above the neutron threshold,
and neutron emission follows the decay.
The neutron emission can affect the {\it r}-process final abundances
by decreasing the neutron number during the decay phase.

In this study, we focus on the impacts of the {\FF} $\beta$~decay
and the $\beta$-delayed neutron emission on the production of the $A \sim 195$ {\it r}-process peak
in several astrophysical models.
We adopt a recent magneto-rotational supernova (MR-SNe) model \cite{2015ApJ...810..109N}
as well as a neutron star merger model \cite{1999ApJ...525L.121F} and
a proto-neutron-star wind (PNSW) model \cite{2007A&A...467.1227A}
that have been used in previous studies.
These astronomical models cover a range of 
realistic physical conditions of the {\it r}-process environments.

\section{Nuclear physics input and astrophysical models}
\label{sec-method}

\subsection{$\beta$~Decay with $\beta$-delayed neutron emission}

\label{sec-fbs}

We adopt $\beta$-decay rates calculated
based on the spherical QRPA method with realistic forces \cite{2013PhRvC..88c4304F},
``FBS13'' hereafter, for nuclei $N \sim 126$.
The QRPA method can handle larger model space than the shell model.
In the region of heavy nuclei,
shell model calculations are usually limited to the $N=126$ isotones,
whereas FBS13 provides theoretical results on $N=124$--$128$ nuclei.
The larger model space also means that negative parity {\FF} transitions were dealt with,
which are missing for some calculations \cite{1995ADNDT..59..185M}.

We apply the rates of $\beta$ decay and $\beta$-delayed neutron emission probabilities
(1n, 2n and 3n emissions) for $N=124$--$128$ nuclei calculated by FBS13.
FBS13 uses experimental $Q_\beta$-values and neutron-separation energies
if available (as in some cases around $N=82$),
otherwise, masses predicted  by the FRDM were employed~\cite{2003PhRvC..67e5802M}.
Due to the difficulty of treating odd-odd nuclei, 
we simply compute their decay rates by means of interpolation from neighbouring nuclei.
Since experiments suggest smooth $\log t_{1/2}$
behaviour for neutron-rich nuclei (see, e.g., \cite{2015PhRvL.114s2501L}),
the properties of odd-odd nuclei were calculated as:
\begin{equation}
t_{1/2}(Z,N)=\sqrt{t_{1/2}(Z+1,N)\times t_{1/2}(Z-1,N)} \ ,
\end{equation}
\begin{equation}
P_{\rm n}(Z,N)=\frac{1}{2} \left[ P_{\rm n}(Z+1,N)+P_{\rm n}(Z-1,N) \right] \ ,
\end{equation}
where $t_{1/2}$ and $P_{\rm n}$ are the $\beta$-decay half-life and neutron emission probability (for 1, 2 and 3 neutron emissions)
for a given nucleus with $Z$ and $N$, respectively.

\begin{figure}[t]
	\begin{center}
    \includegraphics[width=\hsize]{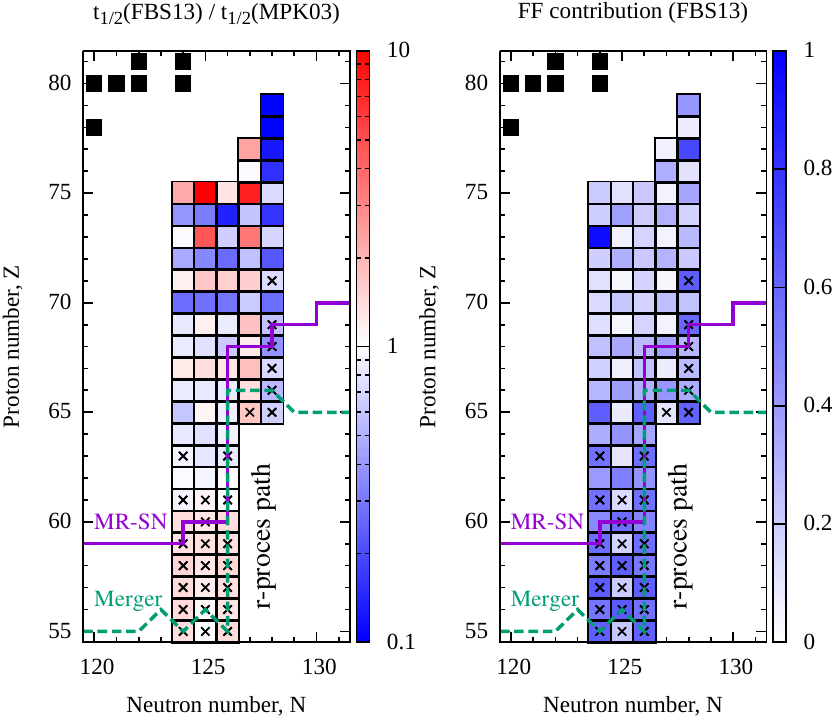}
    \caption{
    The $N$--$Z$ plane for $N \sim 126$;
    Left: the ratio of half-lives, FBS13~\cite{2013PhRvC..88c4304F}
    to MPK03~\cite{2003PhRvC..67e5802M}
    (experimental data are used for near stable nuclei $Z>75$);
    Right: the ratio of {\FF} transition in the total decay rate of FBS13 \cite{2013PhRvC..88c4304F}.    
    In both panels, black filled squares are stable nuclei;
    ``$\times$'' indicates strong neutron emission: $\sum_{n=1}^{3} nP_n > 1$;
    the ``{\it r}-process path'' shows the main path of nucleosynthesis
    of the MR-SN and merger models (see Section~\ref{sec-results}).
    }
  \label{fig-map}
   \end{center}
\end{figure}

For the $N=82$ isotones the adopted half-lives agree well with that of the shell 
model~\cite{2012PhRvC..85a5802S, 2013PhRvC..87b5803Z}
and with the continuum QRPA~\cite{2013PhRvC..88c4304F}.
They are also consistent with the available experimental data.
This is also the case for $N=80$ and $84$ nuclei (here no shell model predictions exist).
In the case of the $N=126$ nuclei, there are no experimental data available.
Here the discrepancy among different calculations is larger than at $N=82$.
The half-life values used here are in agreement with the predictions of the shell model
\cite{2012PhRvC..85a5802S, 2013PhRvC..87b5803Z} within a factor of two.
Regarding the $\beta$-delayed neutron emission probabilities, $P_{\rm n}$,
there are large differences between calculations.
The calculation used here predicts lower values than the shell model 
\cite{2012PhRvC..85a5802S, 2013PhRvC..87b5803Z}
and the continuum QRPA \cite{2013PhRvC..88c4304F} for $N=82$ nuclei.
For $N=126$ nuclei, the differences among calculations are smaller,
all predicting large $P_{\rm n} \sim 0.1$--$1$ values for {\it r}-process path even-even nuclei.
For odd-$Z$ nuclei, small $P_{\rm n}$ values are predicted by all calculations, with the one adopted here predicting lower values
than the others \cite{2012PhRvC..85a5802S, 2013PhRvC..87b5803Z,2013PhRvC..88c4304F}.
As a general rule, more neutron-rich nuclei (lower $Z$) have a larger neutron emission probability.

The {\FF} transitions generally play only a small role for $N=82$ nuclei,
where the ratio of a {\FF} part to the total $\beta$-decay rate is about $10$\%.
However, {\FF} transitions are very important for $N=126$ nuclei,
the contribution of the {\FF} component reaching $20$--$60$\%.
Due to the influence of the {\FF} transition,
the half-lives generally become shorter \cite{2003PhRvC..67e5802M}.

The general trends of the $\beta$-decay for $N \sim 126$ region
are illustrated in Fig.~\ref{fig-map}.
The left panel shows the ratio of half-lives of FBS13
to those of ref. \cite{2003PhRvC..67e5802M} (which will be referred as MPK03 hereafter),
while, the right panel plots the ratio of {\FF} transition to the total decay rate.
Comparing MPK03, FBS13 shows faster decay half-lives in particular for $Z>60$ nuclei.
This is due to the effect of particle-particle residual interactions which brings down the rates,
which are ignored in MPK03 \cite{2013PhRvC..88c4304F}.
In addition, very neutron-rich nuclei,
which have a strong $\beta$-delayed neutron emission $\sum_{n=1}^{3} {n}P_{n} > 1$
(i.e., more than one neutron is emitted per decay),
are marked in both panels.
We will discuss the effects of these decay properties
on the {\it r}-process nucleosynthesis in Section~\ref{sec-results}.

\subsection{Nuclear reaction networks}

\begin{table}
\begin{center}

\caption{Decay rates of networks,
indicating whether $\beta$-decay includes {\FF} transitions and
$\beta$-delayed neutron emission (n.e.) for $N = 124$--$128$ nuclei.
Adopted theoretical rates are based on
MNK97 \cite{1997ADNDT..66..131M}, MPK03 \cite{2003PhRvC..67e5802M}
and FBS13 \cite{2013PhRvC..88c4304F}.}
\begin{tabular*}{\hsize}{@{\extracolsep{\fill}}l|cccc}
\hline
 Network & Base & $N=124$--$128$ & FF & n.e. \\
\hline
MNK97        & MNK97 & MNK97 &--  & -- \\
MPK03        & MPK03 & MPK03 & MPK03 & MPK03 \\
FBS13           & MPK03 & FBS13 & FBS13 & FBS13  \\ 
FBS(w/o FF)  & MPK03 & FBS13 & -- & FBS13 \\
FBS(w/o n.e.) & MPK03 & FBS13 & FBS13 & -- \\
\hline
 \end{tabular*}
\label{tab-network}
\end{center}
\end{table}

We perform nucleosynthesis calculations using a nuclear reaction network code,
described in~\cite{2012PhRvC..85d8801N, 2015ApJ...810..109N, 2012ApJ...758....9N}.
This code consists of more than $4000$ isotopes of $Z \leq 100$.
All nuclear reactions and decay relevant to the {\it r}~process are taken into account,
and fission for heavy nuclei is also included.
For nuclei experimental masses \cite{1995NuPhA.595..409A} are not available,
we adopt theoretical values based on FRDM.
Theoretical rates, e.g., (n,$\gamma$) and $\beta$-decay,
are taken from REACLIB \cite{2000ADNDT..75....1R} and JINA Reaclib \cite{2010ApJS..189..240C}.

Focusing on $\beta$ decay for $N\sim126$ ($N=124$--$128$) nuclei,
we make five different networks summarised in Table~\ref{tab-network}.
Most of theoretical $\beta$-decay rates for thousands neutron-rich nuclei
are taken from ``base'' data, while we modify the decay rate for $N \sim 126$ nuclei.
The first two cases are standard, widely used in nucleosynthesis calculations.
MNK97 (named ``FRDM'' in \cite{2012PhRvC..85d8801N})
includes only $\beta$ decay due to allowed transitions and ignores
$\beta$-delayed neutron emission.
MPK03 includes these effects, calculated with the gross theory of the $\beta$ decay \cite{2003PhRvC..67e5802M}.
FBS13 uses $\beta$~decay half-lives and $\beta$-delayed neutron emission probabilities
calculated for $N=124$--$128$ nuclei \cite{2013PhRvC..88c4304F},
as described in Section \ref{sec-fbs}.
All other nuclear rates are the same as the MPK03 network.

FBS13 is a state-of-the-art calculation focusing on {\FF} transitions
and $\beta$-delayed neutron emission around $N \sim 126$.
Therefore, we adopt these calculations to investigate the effects of these ingredients
on the $A \sim 195$ {\it r}-process abundance peak.
Consequently, we also prepare two additional networks based on FBS.
As listed in Table~\ref{tab-network},
we add FBS(w/o~FF) and  FBS(w/o~n.e.) networks,
in which we ignore the effect of {\FF} in decay rates and neutron emission,
respectively.
These are artificially modified rates for $N \sim 126$ isotones,
where FBS(w/o~FF) has longer half-lives (slower decay) than FBS,
and FBS(w/o~n.e.) simply ignores neutron emission.
Consequently, these additional networks are physically invalid
and will be used only for examining
the role of individual physical processes (in Section~\ref{sec-results}).

\subsection{Astrophysical models}
\label{sec-models}

\begin{figure}[tb]
	\begin{center}
    \includegraphics[width=0.85\hsize]{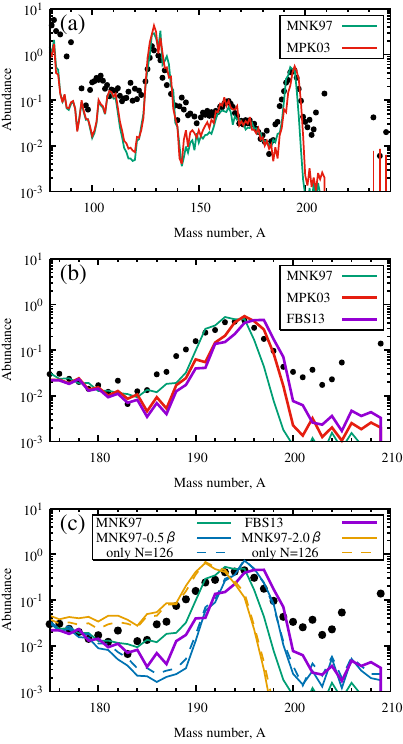}
    \caption{\label{fig-final-abund}
    The final abundances of the MR-SN model for several networks,
    compared with the solar abundances (black dots) \cite {1999ApJ...525..886A}:
    (a) MNK97 and MPK03, covering the entire {\it r}-process range;
    (b) MNK97, MPK03 and FBS13 around the $A=195$ peak;
    (c) MPK03 and its variations around the peak
    (i.e., the solid lines are MNK97-$2.0\beta$ and MNK97-$0.5\beta$,
    of which half-lives of $N=124$--$128$ are varied,
    while dashed-lines are the case $N=126$ isotopes).
    }
	\end{center}
\end{figure}

\begin{figure*}[]
	\begin{center}
    \includegraphics[width=1.0\hsize]{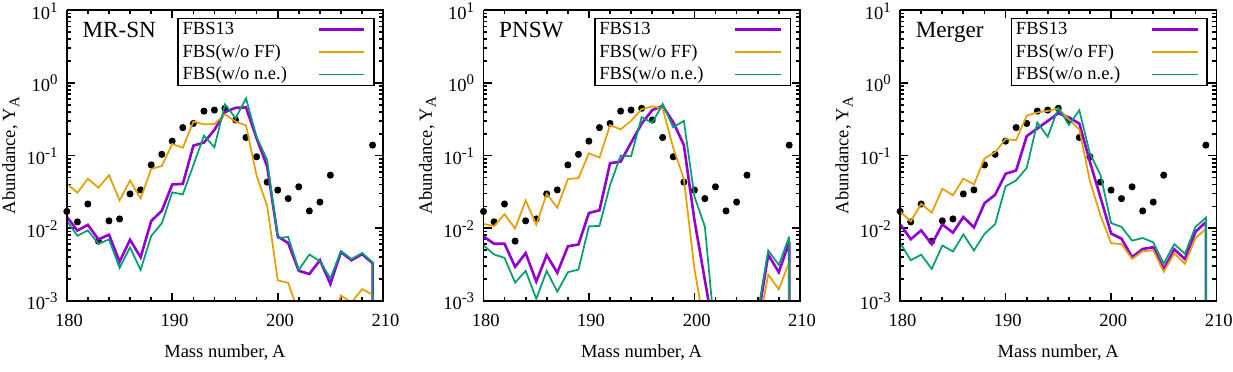}
    \caption{\label{fig-comp-all}
    The final abundances of the MR-SN, the PNSW, and the NS--SN merger models.
    The final abundances based on networks of FBS13, FBS(w/o~FF) and, FBS(w/o~n.e.)
    are shown together with the solar {\it r}-process abundances (the black dots \cite {1999ApJ...525..886A}).}
	\end{center}
\end{figure*}

We calculated the {\it r}~process nucleosynthesis
based on several astrophysical conditions
(i.e., the density, temperature and electron fraction $Y_{\rm e}$ or initial composition).
We adopted the jet-like explosion of magneto-rotational supernovae (MR-SNe) as a fiducial case,
providing medium neutron-rich ($Y_{\rm e} \sim$ $0.1$--$0.3$) environments
and relatively low entropies $\sim 10~k_{\rm B}~{\rm baryon}^{-1}$.
We use an energetic jet-like explosion model with strong effects of rotation and magnetic fields,
classified as ``prompt-magnetic-jets'' \cite{2009ApJ...691.1360T}.
The details of the nucleosynthesis are described in~\cite{2015ApJ...810..109N},
and the role in the chemical evolution is discussed in \cite{2015ApJ...811L..10T}.

We also calculate nucleosynthesis for a proto-neutron star wind (PNSW) model,
which is slightly neutron-rich with significantly higher entropies $\sim 100 k_{\rm B}~{\rm baryon}^{-1}$.
We adopt a trajectory from 1D hydrodynamical calculations
of $15 M_\odot$ progenitor (model {\tt r15-l1-r1} of \cite{2007A&A...467.1227A}).
We assume $Y_{\rm e} = 0.3$ with an original hydrodynamical evolution of the temperature
and density ($S \sim 165~k_{\rm B}$).
In addition, we also calculate {\it r} process using the trajectory of a neutron star merger (NS--NS) model,
based on a hydrodynamical simulation \cite{1999ApJ...525L.121F}.
The physical properties of ejecta for NS--NS mergers is still unsettled,
where previous simulations have shown extremely neutron-rich $Y_{\rm e}< 0.1$ environments
with strong effect of fission-cycling.
We assume that the main ejecta has $Y_{\rm e} > 0.1$,
based on a more recent treatment of the ejecta by shock-heated ejection mechanism
\cite{2014ApJ...789L..39W, 2015MNRAS.452.3894G},
where weak interaction is activated,
and the $Y_{\rm e}$ is considerably higher than $0.1$.
We select a trajectory with $Y_{\rm e} = 0.14$
that produces {\it r}-process nuclei including the $N=126$ peak.

\section{Results}
\label{sec-results}

We performed nucleosynthesis calculations using different $\beta$-decay rates for the $N \sim 126$ nuclei,
as summarised in Table~\ref{tab-network}.
The abundance of the MR-SN model based on the two standard nuclear physics inputs
are shown in Fig.~\ref{fig-final-abund}(a). 
The global features of abundance
patterns in the entire r-process elements are similar, and the abundances
of $A < 180$ nuclei are almost identical.
However, the abundances around the {\it r}-process third peak
at $A \sim 195$ are different. 
Fig.~\ref{fig-final-abund}(b) focuses on this $A \sim 195$ abundance peak.
The results of FBS13 are also included.

The peak of FBS13 shifts to the higher mass number
and it is wider compared with both MPK03 and MNK97 which are based on FRDM calculations.
The shift of the peak is simply understood by the difference of $\beta$-decay timescales.
Shorter decay half-lives aid the production of heavier nuclei,
because the $\beta$ decay determines the progress of the {\it r} process
around the waiting point nuclei
(as already investigated \cite[][]{2003PhRvC..67e5802M,
2001PhRvC..64c5801S, 2014arXiv1405.0210C, 2015ApJ...808...30E}).
MPK03 has shorter half-lives than MNK97 owing to the inclusion of {\FF} transitions.
FBS13 also includes {\FF} transitions and its half-lives are generally shorter
than those of MPK03 for the r-process path nuclei due to the effects of particle-particle residual interactions
\cite{2013PhRvC..88c4304F} (see also the left panel of Fig.~\ref{fig-map}).

The impact of half-lives for $N\sim126$ also clearly appears in Fig.~\ref{fig-final-abund}(c)
by the comparison of MNK97 and its variation.
MNK97-$0.5\beta$ has shorter half-lives multiplied by $0.5$ and
MNK97-$2.0\beta$ has longer half-lives by $2.0$, respectively.
For solid lines, we artificially change the half-lives for $N=124$--$128$ nuclei
and we only modified $N=126$ isotones for dashed lines.
Here, abundances of each solid line are similar to the ones of corresponding dashed line.
This is understood as the ``{\it r}-process path'' in Fig.~\ref{fig-map} is mostly on the $N=126$,
and the half-lives of $N=126$ isotones
are dominant for determining the progress of the $\it r$ process.

\begin{figure*}[t]
	\begin{center}
    \includegraphics[width=1.0\hsize]{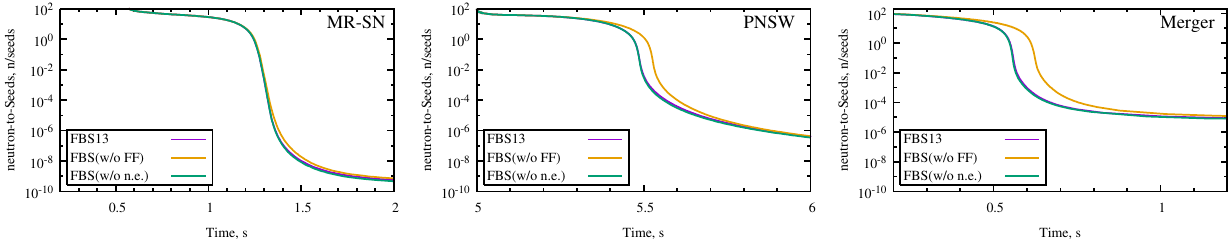}
    \caption{\label{fig-nseed}
    The evolution of the neutron-to-seed ratio (n/seed) of MR-SN, PNSW and NS--NS merger models
    for three different networks of FBSs.}
	\end{center}
\end{figure*}

We examine the effects of individual physical processes,
i.e., the {\FF}-transitions to $\beta$~decay and neutron emission
by comparing the reaction network of FBS13 and its sub-networks.
The results of the calculations are shown in Fig.~\ref{fig-comp-all}
for the three astrophysical models (i.e., the MR-SN, PNSW and NS merger).
The FBS(w/o~FF) network uses longer $\beta$-decay half-lives,
as we ignored the decay components due to the {\FF} transitions.
In all the cases, the peak of FBS13 is located at larger mass number
compared with the FBS(w/o~FF).
This feature is valid to all models, although the merger model
has smaller impacts in $A>195$ region than MR-SN and PNSW.
This is due to the strong progress of neutron captures with fast expansion,
which locates the {\it r}-process path at very neutron-rich region.
Shifting the peak to higher mass numbers is naturally understood
by the shorter half-lives of FBS13 compared with FBS(w/o~FF).
This is consistent with the results in Fig.~\ref{fig-final-abund},
that shorter $\beta$-decay half-lives support production of heavier elements.

The other sub-network FBS(w/o~n.e.) uses the same decay rates (half-lives) as FBS13,
but we artificially ignored the effects of $\beta$-delayed neutron emission.
We find two characteristics in the abundance patterns:
(i) the inclusion of neutron emission shifts the lower end of the $A \sim 195$ abundance peak to lower mass region, resulting in a broader peak in all three astrophysical models;
(ii) neutron emission washes out the odd-even mass staggering and results in a smoother abundance peak.

Nuclei with high neutron emission numbers (more than one neutron emitted on average,
$\sum_{n=1}^{3} nP_n > 1$)
are marked by ``$\times$'' in Fig.~\ref{fig-map}.
Nuclei on the {\it r}-process path have significant neutron emission probability
and they are higher for nuclei further from stability (e.g., lower $Z$ with $N=126$). 
Consequently neutron emission widens the final abundance pattern in the direction of lower $A$.
This widening is dependent on the astrophysical model. 
Models in which the {\it r}-process path passes through higher $Z$ nuclei, such as the MR-SN (see Fig.~\ref{fig-map}),
the higher mass end of the abundance peak is hardly affected,
and only the lower end shifts to lower masses.
This causes a significant broadening of the peak.
In contrast, in the merger model the {\it r}-process path lies further from stability,
and the neutron emission moves the whole abundance peak to lower $A$.
We note that the strength of neutron emission is higher for odd $Z$ nuclei than for even
$Z$ ones (for more details, see \cite{2013PhRvC..88c4304F}).
This asymmetry changes the path of the decay flow
and the even-odd mass scattering in the abundance peak.

Finally, we examine the role of $\beta$ delayed neutrons and {\FF} transitions
by describing the time evolution of the neutron density.
The temporal evolution of the neutron-to-seed ratio (n/seed, the ratio of the neutron number density to the seed nuclei)
is shown in Fig.~\ref{fig-nseed}.
The choice of networks and astrophysical models is the same as Fig.~\ref{fig-comp-all}.
In all three astrophysical scenarios, the inclusion of the {\FF} transitions
(shorter $\beta$-decay half-lives) reduces the time window with 
large n/seed, therefore reducing the timescale of the neutron capture phase of the {\it r}~process.
The effect is the largest for the merger and smallest for the MR-SN model.
On the other hand, the $\beta$-delayed neutron emission slightly increases the n/seed.
The increase is, however, small to affect heavy element production.
The $\beta$-delayed neutron emission mostly has impacts
to form the $N \sim 126$ peak during $\beta$-decay phase,
as shown in Fig.~\ref{fig-comp-all}.

\section{Conclusion}
\label{sec-conclusion}

We studied the role of first-forbidden $\beta$~decay and that of the $\beta$-delayed neutron emission on the {\it r} process,
specifically the effect on the $A \sim 195$ peak.
We used the state-of-the-art calculations of $\beta$-decay rates and neutron emission probabilities \cite{2013PhRvC..88c4304F}
based on three astrophysical scenarios.
The reached conclusions are general, and are independent of the chosen nuclear physics input calculation.
We found the following impacts on the production of the $A=195$ {\it r}-process abundance peak:
\begin{itemize}
\item 
The inclusion of the first-forbidden $\beta$~decay
leads to faster progress of the {\it r}~process for the waiting-point nuclei.
This is mostly caused by the shorter $\beta$-decay half-lives of the $N \sim 126$ nuclei.
This shifts the $A \sim 195$ peak towards higher masses and results in slightly wider distribution.
\item The delayed neutron emission widens the abundance peak in the direction of smaller neutron number, modifying its lower mass tail.
The asymmetry of neutron emission probability between the odd and even $Z$ nuclei 
smoothens the final abundances during the decay phase.
\end{itemize}
These characteristics are commonly seen in the all adopted astrophysical models
(i.e., MR-SN, NS--NS merger and PNSW models),
although the size of the effects depend on physical conditions and details of the nuclear physics inputs.

The theoretical uncertainty of $\beta$-decay properties is still large,
as the relevant nuclei in the $N =126$ region are presently beyond the reach of experiments.
For example a recent calculation in the Ni region \cite{2015PhRvL.114n2501N} implies
that the particle-vibration coupling effect beyond QRPA could shift
the Gamow-Teller strength downwards energetically,
which increases the decay rates and reduces the ratios of the FF part of the total decay.
In the future, improved evaluations of the ratios of FF to Gamow-Teller contributions
would take into account these effects beyond QRPA.
The production of the $A \sim 195$ peak
is also affected by fission including neutron emission \cite{2015ApJ...808...30E},
as well as $\beta$-decay half-lives.
In order to reduce uncertainty in nucleosynthesis calculations,
complete knowledge about decay properties for very neutron-rich nuclei is essential.
We absolutely need experimental information on the waiting point nuclei.
Experiments for the measurement of $N = 126$ nuclei are progressing,
with plans to produce these nuclei in both deep-inelastic \cite{2013NIMPB.317..752W, 2015PhRvL.115.172503W}
and fragmentation reactions \cite{2008NIMPB.266.4589P}.

\section*{Acknowledgements}

We thank the referee for her/his fruitful comments to improve the manuscript.
We also acknowledge B.~A.~Brown for $\beta$-decay rates,
and A.~Arcones, T.~Kajino, H.~Miyatake, T.~Takiwaki and F.-K.~Thielemann
for discussion and/or for providing numerical data of astrophysical models.
N.N. and Zs.P. were supported by BRIDGCE UK Network (www.bridgce.ac.uk).
The work has been supported by European Research Council (EU-FP7-ERC-2012-St Grant 306901),
by STFC (UK) and by Grants-in-Aid for Scientific Research (C) 15K05090 of the MEXT of Japan.
N.N. carried out numerical calculations on computers at Center for Computational Astrophysics,
National Astronomical Observatory of Japan.

\end{document}